\documentclass[a4paper]{jpconf}
\usepackage{graphicx}
\usepackage{amsfonts}
\usepackage{amsmath}
\usepackage{amssymb}
\usepackage{graphics,graphicx}
\usepackage{epsf}
\usepackage[%
  colorlinks=true,
  urlcolor=blue,
  linkcolor=blue,
  citecolor=blue
]{hyperref}
\usepackage[all]{hypcap}
\usepackage{color}
\usepackage[utf8]{inputenc}

\begin{document}

\title{DMRG study of FQHE systems in the open cylinder geometry}

\author{G. Misguich and Th. Jolicoeur}

\address{Universit\'e Paris-saclay, CNRS, CEA, Institut de Physique Th\'eorique, Gif sur Yvette, 91190 France}

\ead{gregoire.misguich@ipht.fr,thierry.jolicoeur@ipht.fr}

\begin{abstract}
The study of the fractional quantum Hall liquid state of two-dimensional electrons requires a non-perturbative treatment of interactions. It is possible to perform exact diagonalizations of the Hamiltonian provided one considers only a small number of electrons
in an appropriate geometry. Many insights have been obtained in the past from considering electrons moving on a sphere or on a torus. In the Landau gauge it is also natural to impose 
periodic boundary conditions in only one direction, the cylinder geometry. The interacting 
problem now looks formally like a one-dimensional problem that can be attacked by the standard DMRG algorithm. We have studied the efficiency of this algorithm to study the ground state properties of the electron liquid at lowest Landau level filling factor $\nu=1/3$ when the interactions are truncated to the two most important repulsive hard-core
components. Use of finite-size DMRG allows us to conclude that the ground state is a compressible two-electron bubble phase in agreement with previous Hartree-Fock calculations. We discuss the treatment of Coulomb interactions in the cylinder geometry. To regularize the long-distance behavior
of the Coulomb potential, we compare two methods~: using a Yukawa potential or forbidding
arbitrary long distances by defining the interelectron distance as the \textit{chord} distance through the cylinder. This allows us to observe the Wigner crystal state for small filling factor.
\end{abstract}

\section{Introduction}

When two-dimensional electrons are submitted to a magnetic field perpendicular to the plane of
allowed motion their one-body wavefunctions are the so-called Landau orbitals that form a
regularly spaced ladder of states, the Landau levels, and each member of the ladder has the
same degeneracy given by the ratio of the magnetic flux through the sample and the elementary
flux quantum $N_\Phi=\Phi/\Phi_0$. When the electron density is such that an integer number of these
levels is filled, we are in the conditions of the integer quantum Hall effect and the physics
of this situation can be understood essentially without considering interactions. However when there is partial filling the electrons form a remarkable state of matter, incompressible liquids with excitations possessing fractional charge as well as fractional statistics. 
This is the fractional quantum Hall effect (FQHE).
These liquids appear for special rational values of the filling factor $\nu=N_e/N_\Phi$.
The most prominent state appears in the lowest Landau level for $\nu=1/3$. From a theoretical point of view we are facing a tough problem because the kinetic energy is frozen in a Landau level and the fate of electrons is decided entirely by the two-body Coulomb interactions. So
no perturbative approach is feasible. One tool that has proven very fruitful is exact
diagonalization (ED) of systems of few electrons when the Fock space is small enough to store
a handful of vectors in computer memory allowing use of Krylov subspace methods. However the limitations on the number of electrons is too severe to consider problems involving for example additional degeneracies like spin or valley degrees of freedom that are very important in
the physics of two-dimensional materials like graphene. It has long been noted that since
the Landau gauge orbitals can be naturally indexed by an integer, the projected Coulomb interaction
is formally a one-dimensional problem albeit with long-range interactions.
It is thus possible to use the density matrix renormalization group 
algorithm~\cite{white_density_1992,SchollwockAnnPhys11,itensor3}
 to FQHE
provided one chooses the correct geometry. Recent progresses have been 
done in this direction~\cite{shibata_ground-state_2001,yoshioka_dmrg_2002,shibata_ground_2003,feiguin_density_2008,kovrizhin_density_2010,zhao_fractional_2011,zaletel_infinite_2015,zhu_fractional_2015,zhu_topological_2015,johri_probing_2016,zhu_fractional_2016}.
here we present the use of DMRG in the geometry of a cylinder with open boundaries which has been explored already by ED. We made use of the modern C++ library ITensor (version 3.1)~\cite{itensor3}. The DMRG gives easy access to
an approximation of the ground state wavefunction from which one can compute observables. We take profit of this feature to obtain the pair correlation function, a standard diagnostic tools for the
properties of quantum liquids.

The section (\ref{defs}) defines what we call the cylinder geometry with open boundary conditions.
In section (\ref{hard}) we present a study of a special truncated interaction between electrons.
We investigate the case of the Coulomb interaction in section (\ref{CoulombRes}).
Finally section (\ref{conclude}) contains our conclusions.

\section{The open cylinder geometry for FQHE}
\label{defs}

In the Landau gauge the one-body eigenstates spanning 
the lowest Landau level are given by~:
\begin{eqnarray}
\phi_n(x,y)=\left(\frac{1}{L l \sqrt{\pi}}\right)^{\frac{1}{2}} \exp\left(-\frac{1}{2l^2}(x-x_n)^2\right)   \exp\left(i k_n y\right),
\end{eqnarray}
where~:
\begin{equation}
 k_n=\frac{2\pi n}{L},\quad x_n/l= - \frac{2\pi nl}{L}.
\end{equation}
Here we have imposed periodic boundary conditions along the $y$ direction $\phi(y)=\phi(y+L)$, leading to quantization of momentum $k_n$. The integer $n$ can be positive or negative and defines also
the center of the Gaussian wavepacket in the $x$ direction.
We have used $l=\sqrt{\hbar/eB}$ which is the magnetic length, set to unity in the rest of the paper.
In the following we consider \textit{finite} cylinders, obtained by considering a finite number $N_{\rm orb}$ of orbitals. So we truncate the Hilbert space,
constraining the integer $n$ to take only $N_{\rm orb}$ different values. It is important to note that while such a truncation
is mathematically convenient it is not produced by imposing some physical hard-wall
condition since the one-body eigenstates have a Gaussian shape in $x$ so formally extend
far away (but with fast decay).
The model does therefore not have a sharp boundary in real space.
Since the spacing in the $x$ direction between two consecutive orbitals is $\delta x= 2\pi/L$, the length in $x$ of the cylinder
is of order $L_x\simeq N_{\rm orb} \delta x = 2\pi N_{\rm orb} /L$.
For an odd number $N_{\rm orb}$ of orbitals we restrict the index $n$ to be in $-(N_{\rm orb}-1)/2 \leq n \leq (N_{\rm orb}-1)/2$, and for even $N_{\rm orb}$ we take $-N_{\rm orb}/2+1 \leq n \leq N_{\rm orb}/2$. In the following we will denote by $\mathcal{I}$
this set of integers. So we see that there are two independent parameters ruling the spatial
extent in this geometry~: the perimeter $L$ of the cylinder and the $x$ extent of the region
where electrons can roam. Fixing the electron number $N_e$ and the filling factor means that we 
fix the number of orbitals and thus $x$ extent, the only remaining parameter is then the length $L$.
It is intuitively obvious that this length should be scaled with $N_e$ as the numbers of particles and orbitals grow, so that the region
allowed to electron motion is more or less square~\cite{Soule1,Soule2,Soule3}. Strong deviations from this case leads
to behavior unrelated to the thermodynamic limit of a two-dimensional system. For example
when $L\rightarrow 0$, the ``thin torus'' limit, one is led to an electrostatic problem.  In the opposite ``hoop'' limit, $L\rightarrow\infty$, one collapses the system into a one-dimensional Luttinger liquid with no remnant of the bulk.

The second-quantized formula of a generic two-body Hamiltonian is given by~:
\begin{equation}
 \mathcal{H}= \frac{1}{2}\sum_{n_1 n_2 n_3 n_4} \mathcal{A}_{n_1 n_2 n_3 n_4} c_{n_1}^\dag c_{n_2}^\dag c_{n_3} c_{n_4}
 \label{eq:H}
\end{equation}
where the matrix elements are related to the real space potential $V$ through
\begin{align}
 \mathcal{A}_{n_1 n_2 n_3 n_4}= \int d{\bf r}_1 d{\bf r}_2 \,\,\phi_{n_1}( {\bf r}_1)^*\phi_{n_2}( {\bf r}_2)^* 
  V({\bf r}_1-{\bf r}_2) \phi_{n_3}( {\bf r}_2)\phi_{n_4}( {\bf r}_1). 
  \label{eq:Adef}
\end{align}
To evaluate these matrix elements it is convenient to go to momentum space~:
\begin{eqnarray}
 \mathcal{A}_{n_1 n_2 n_3 n_4}
 &=& \int \frac{d{\bf Q}}{(2\pi)^2}
 F_{n_1,n_4}({\bf Q}) F_{n_2,n_3}(-{\bf Q})\tilde V({\bf Q}),
 \label{eq:Ageneral}
\end{eqnarray}
where we have defined~:
\begin{equation}
 F_{n,m}({\bf Q})=\int d{\bf r}\,\, \phi_{n}( {\bf r})^*\phi_{m}( {\bf r})
 e^{-i{\bf Q}\cdot{\bf r}}
\end{equation}
and the Fourier transform~:
\begin{equation}
 \tilde V({\bf Q})=\int d{\bf R} \,\, V({\bf R}) e^{i{\bf Q}\cdot{\bf R}}.
\end{equation}
Due to the periodicity of the wavefunction and of the interaction in the $y$ direction we only need to consider wave-vectors ${\bf Q}$ of the form ${\bf Q}=(q_x,q_y=2\pi q /L)$ with $q\in\mathbb{Z}$. In such cases the function $F$ is equal to~:
\begin{eqnarray}
 F_{n,m}({\bf Q}=
 (q_x,q_y=2\pi q /L)) = \delta_{q,m-n}
 \exp\left(
 -\frac{1}{4}q_x^2+\frac{i}{2}q_x\left(k_n+k_m\right)-\frac{1}{4}\left(k_n-k_m\right)^2
 \right)
 \label{eq:Fnm}
\end{eqnarray}
with $k_i={2\pi n_i}/{L}$.

In the following we will consider two versions of the  $1/r$ Coulomb interaction which are adapted to the cylindrical geometry. Our use of $1/r$ means that energies are measured in units of
$e^2/\epsilon l$ with $e$ the electric charge and $\epsilon$ the dielectric constant of the host
material.

\subsection{Yukawa}

One standard way to make the Coulomb interaction periodic in the $y$ direction of the cylinder ($V(x,y)=V(x,y+L)$) is  to use a sum over periodic
images~:
\begin{equation}
 V({\bf R})=\sum_{n=-\infty}^\infty \frac{1}{\left|{\bf R}+Ln{\bf e}_y \right|}.
 \label{eq:coulomb_images}
\end{equation}
In Fourier space this is equivalent to~:
\begin{equation}
 \tilde V({\bf Q}=(q_x,q_y))=\frac{2\pi}{\left|{\bf Q}\right|} \sum_{n=-\infty}^\infty e^{i n L q_y}.
\end{equation}
The above sum of exponentials is proportional to a Dirac ``comb'' (Poisson formula) and we obtain~:
\begin{equation}
 \tilde V({\bf Q})=\frac{(2\pi)^2}{L \left|{\bf Q}\right|}
 \sum_{n=-\infty}^\infty \delta\left(q_y-\frac{2\pi n}{L}\right).
\end{equation}
One possibility to regularize the long-distance tail of the Coulomb interaction is by introducing some (small) Yukawa mass $\mu$. The main effect is
to replace the Coulomb interaction by an exponential decay beyond the length scale $\mu^{-1}$.
In that case the Fourier transform of the potential becomes~:
\begin{equation}
 \tilde V({\bf Q})=\frac{(2\pi)^2}{L \sqrt{q_x^2+q_y^2+\mu^2}} \sum_{n=-\infty}^\infty \delta\left(q_y-\frac{2\pi n}{L}\right).\label{eq:VQyuka}
\end{equation}
This leads to an expression for the matrix elements~:
\begin{eqnarray}
 \mathcal{A}_{n_1 n_2 n_3 n_4}=\int_{-\infty}^{\infty} \frac{dq_x}{L}
 \sum_{n}
 \frac{1}{\sqrt{q_x^2+\left(\frac{2\pi n}{L}\right)^2+\mu^2}} 
F_{n_1,n_4}(q_x,\frac{2\pi n}{L}) F_{n_2,n_3}(-q_x,-\frac{2\pi n}{L} ).
 \label{eq:Ayuka1}
\end{eqnarray}

Taking into account translation invariance we have $k_1-k_3=k_4-k_2$ and we finally obtain~:
\begin{eqnarray}
\mathcal{A}_{n_1 n_2 n_3 n_4}=
\frac{2 \delta_{n_1+n_2,n_3+n_4}}{L}
\exp\left(-\frac{1}{2}\left(k_1-k_4\right)^2\right)
\int_{0}^{\infty}  
\frac{dq_x \cos\left(q_x\left(k_1-k_3\right)\right)}{\sqrt{q_x^2+\left(k_1-k_4\right)^2+\mu^2}} 
\times\exp\left(-\frac{1}{2}q_x^2 \right).
\end{eqnarray}
To obtain the matrix elements in the simulations the above integrals have to be computed numerically.
Note also that for $n_1=n_4$ (and thus $n_2=n_3$) the integral is infra-red divergent when $\mu\to0$
and we have $\mathcal{A}_{n,m,m,n}\sim |\log(\mu)|$.

\subsection{Coulomb-chord potential}
The disadvantage of the Yukawa potential is the
presence of an extra parameter and the need
to extrapolate the physical observables to $\mu\to0$,
in addition to taking thermodynamic limit ($N_{\rm orb},N_e\to\infty$).
We now turn to another formulation of the Coulomb interaction on a cylinder.
To make $V({\bf R})$ periodic in the $y$ direction of the cylinder, the Euclidean distance
$d^2=R_x^2+R_y^2$ is replaced by the chord distance~:
\begin{equation}
 d^2=R_x^2+\left(\frac{L}{\pi}\right)^2 \sin^2\left(\frac{\pi R_y }{L}\right).
\end{equation}
We will now proceed to the calculation of the associated matrix elements $\mathcal{A}_{n_1,\cdots,n_4}$.
To do so it is convenient to use an integral representation of $1/d$~:
\begin{equation}
 V({\bf R})=1/d=\int_{-\infty}^{\infty} \frac{d\alpha}{\sqrt{\pi}}\exp{\left(-\alpha^2 d^2\right)}.
\end{equation}
In Fourier space we have~:
\begin{eqnarray}
 \tilde V({\bf Q}) =
 \int d{\bf R}
 \int_{-\infty}^{\infty} \frac{d\alpha}{\sqrt{\pi}}
 \exp\left(-i q_x R_x-i q_y R_y
 -\alpha^2 \left[R_x^2+\left\{\frac{L}{\pi}
 \sin\left(\frac{\pi R_y}{L}\right)\right\}^2\right]\right).
 \end{eqnarray}
After a few simplifications (including a Gaussian integration over $R_x$ and then over $q_x$) this can be finally re-expressed as~:
\begin{eqnarray}
 \mathcal{A}_{n_1 n_2 n_3 n_4}= \frac{4}{\pi^{3/2}}
e^{-\frac{1}{2}(k_1-k_4)^2}
 \int_{0}^{\pi/2} dv\int_0^\infty \frac{d\alpha}{\sqrt{2\alpha^2+1}}
 \cos\left(vN\right)
 e^{
 -\frac{\alpha^2(k_1-k_3)^2}{2\alpha^2+1}
 -\left\{ \frac{\alpha L}{\pi}\sin(v)\right\}^2
 }. 
 \end{eqnarray} 
 with $N=n_4-n_1+n_2-n_3$ an even integer.

\subsection{Neutralizing background}

In the absence of a neutralizing background, and due to the long-range nature of the Coulomb interaction, the particles will have a strong tendency to accumulate at the edges
of the cylinder. This effect is compensated by a neutralizing background.
We consider the following electric charge density (of a sign opposite to that of the electrons):
\begin{equation}
 \rho_{\rm backg.}({\bf r})=-\nu \sum_{n\in\mathcal{I}} |\phi_n({\bf r})|^2
\end{equation}
where $\nu=N_e/N_{\rm orb}$. This charge density is not strictly uniform, but it becomes uniform in the limit of large cylinder circumference $L$.
The background term in the Hamiltonian is then~:
\begin{equation}
 \mathcal{H}_B = \int d{\bf r} d{\bf r'} \rho_{\rm backg.}({\bf r}) V({\bf r}-{\bf r}') \hat\psi^\dag ({\bf r'}) \hat\psi ({\bf r'})
\end{equation}
with $\hat\psi({\bf r'})=\sum_n  \phi_n({\bf r'})c_n$.
It is also equal to~:
\begin{eqnarray}
 \mathcal{H}_B = -\nu \sum_{n,m\in\mathcal{I}} A_{n,m,m,n} c^\dag_n c_n
 = \sum_n B_n c^\dag_n c_n \;\;{\rm with}\;\;B_n=-\nu \sum_m A_{n,m,m,n}.
\end{eqnarray}

There is also the electrostatic interaction energy of the background with itself~:
\begin{eqnarray}
 \mathcal{E}_{BB}= \frac{1}{2 }\int d{\bf r} d{\bf r'} \rho_{\rm backg.}({\bf r}) V({\bf r}-{\bf r}') \rho_{\rm backg.}({\bf r'})
=\frac{\nu^2}{2} \sum_{n,m\in\mathcal{I}} A_{n,m,m,n}.
 \end{eqnarray}
Since we work at fixed number of fermions, $\sum_n c^\dag_n c_n = N_e$, the above constant can be absorbed into a uniform shift of the background potential
$ B_n \to B_n + \mathcal{E}_{BB}/N_e$.

\subsection{Measurements}
Simple observable quantities include the local density~:
\begin{eqnarray}
 \rho({\bf r})=\sum_{n\in\mathcal{I}} | \phi_n({\bf r})|^2  \langle c^\dag_n c_{n}\rangle
   =\frac{1}{Ll\sqrt{\pi}}\sum_{n\in\mathcal{I}}\exp\left(\frac{(x-x_n)^2}{l^2}\right)  \langle c^\dag_n c_{n}\rangle.
\end{eqnarray}
Another quantity of interest is the pair correlation function~:
\begin{equation}
\label{eq:G_def}
 G({\bf r}_1,{\bf r}_2)=\sum_{i\ne j}\langle \delta({\bf r}_1-\hat {\bf r}_i) \delta({\bf r}_2-\hat {\bf r}_j)        \rangle,
\end{equation}
which can be expressed in second quantization as~:
\begin{eqnarray}
 G({\bf r}_1,{\bf r}_2) = \langle  \hat\psi^\dag ({\bf r}_1) \hat\psi^\dag ({\bf r}_2) \hat\psi ({\bf r}_2)   \hat\psi ({\bf r}_2)   \rangle 
  =
  \sum_{i,j,k,l\in\mathcal{I}}
    \phi^*_i ({\bf r}_1) \phi^*_j ({\bf r}_2) \phi_k ({\bf r}_2)\psi_l ({\bf r}_1) \langle  c^\dag_i c^\dag_j  c_k c_l     \rangle.
\end{eqnarray}

\section{The hard-core model for spinless fermions at filling $\nu=1/3$}
\label{hard}

When projected onto the lowest Landau level, any two-body interaction can be parametrized
by a discrete set of energies called the Haldane pseudopotentials. They are the exact energies of the 
two-body problem which is solvable in this special case. We call them $V_m$
and they are indexed by the relative angular momentum $m$ of the two-particle problem which is a positive
integer. For spinless fermions, due to the Pauli principle only odd values matter.
So a generic two-body Hamiltonian can be written as~:
\begin{align}
\mathcal{H}= \frac{(2\pi)^{5/2}}{L^3}\sum_k V_k\sum_{\{n_i\}} \lambda^{(n_1-n_4)^2+(n_1-n_4)^2}\,\,
H_k((n_1-n_2)/R)H_k((n_4-n_3)/R)\,\,
c_{n_1}^\dagger c_{n_2}^\dagger c_{n_3}c_{n_4}
\label{pseudoH}
\end{align}
where the first sum is over odd integers $k$ and $H_k$ are Hermite orthogonal polynomials.
We have introduced $\lambda=\exp(-2\pi^2/L^2)$ as well as the radius of the cylinder $L=2\pi R$. The thermodynamic limit requires that
we go to large $L$ hence $\lambda$ should be close to unity.
If we drop all $V_k$ except $V_1$ one obtains a model whose ground state for filling factor $\nu=1/3$
is exactly given by the Laughlin wavefunction. The realistic Coulomb potential has all nonzero
pseudopotentials that are decreasing with increasing $k$.

\begin{figure}
\centering
 \includegraphics[width=0.9\columnwidth]{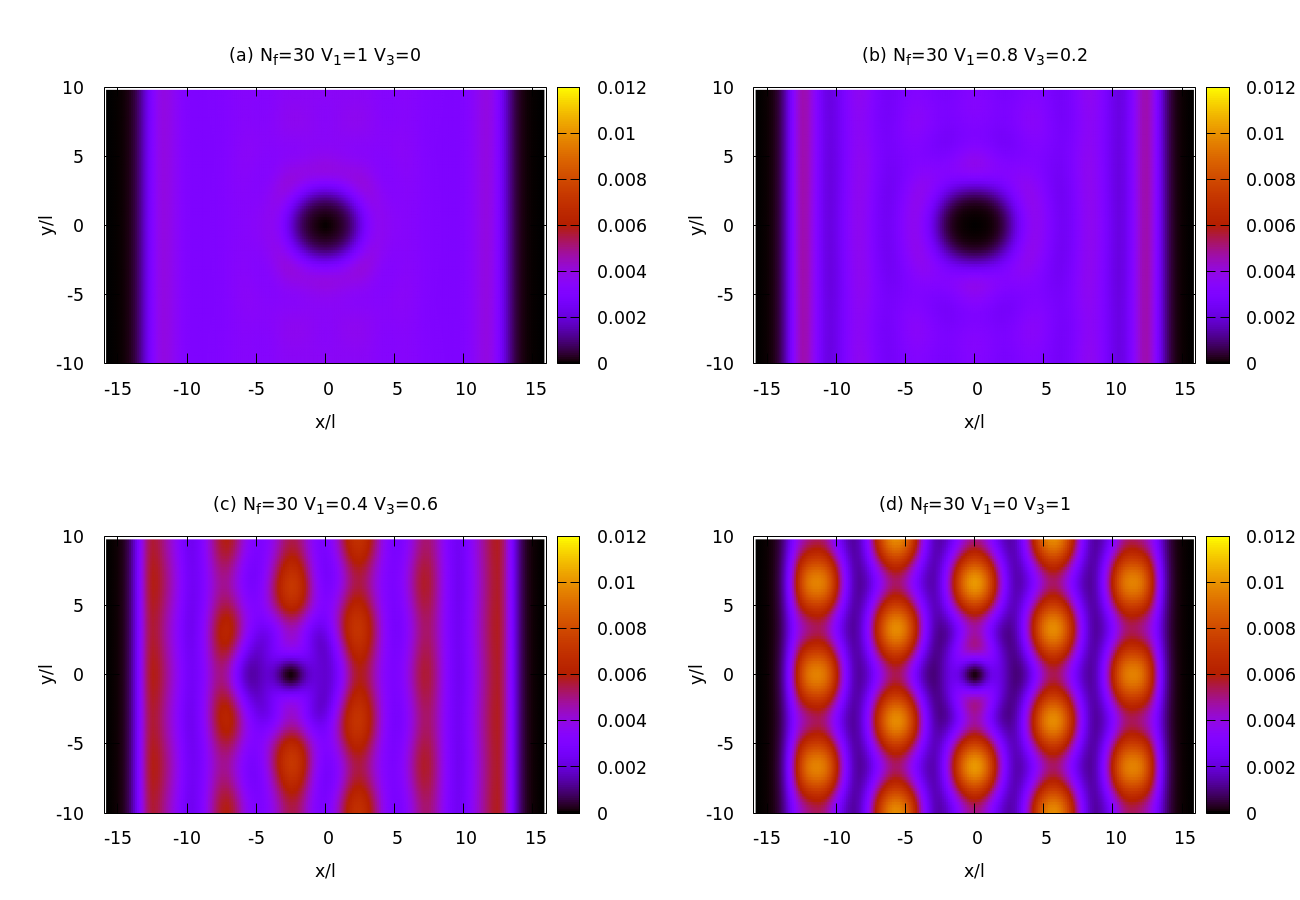}
 \caption{The pair correlation function $g(r)$ computed in real space for various models interpolating between pure $V_1$ and pure $V_3$ models. While the pure $V_1$ case is a featureless liquid, density modulations develop with increasing $V_3$. In panel (d) we observe formation of
 a two-dimensional structure~: a triangular crystal of overdensities containing exactly two
 electrons per site. In panel (c) the reference electron has been offset from the center for
 clarity. In all cases the largest bond dimension of the matrix-product state is of the order of 8000 and the largest discarded weight varies between $ \sim 6.10^{-6}$ (for $V_1=0.4$) and $\sim4.10^{-8}$ (for $V_1=0.8$).}
 \label{bubble}
\end{figure}

Having the Laughlin wavefunction as the unique ground state of the pure $V_1$ Hamiltonian requires
a special tuning of the number of orbitals versus the number of particles~: $N_{\rm orb}=3(N_e-1)+1$.
The constant term in this relation, dubbed ``shift'', has a topological significance. It has been proposed~\cite{Wojs2004} that if we
consider the model Hamiltonian with now only $V_3$ pseudopotential then another incompressible state
is obtained for $N_{\rm orb}=3N_e-6$. This relation leads to the same filling factor $\nu=1/3$
as the Laughlin state but with a different shift.
This proposal was based on limited evidence from ED studies in the spherical geometry. Such a state if confirmed would be a new class of topological order different from the well-studied Laughlin state. We have undertaken a DMRG study of the generalized model including both $V_1$ and $V_3$ pseudopotentials to search for some change of the physical properties when interpolating between these 
two special Hamiltonians~\cite{hardcore-us}. A simple probe of quantum liquids is the pair correlation function $g({\bf r})$.
We compute it for systems of up to 30 fermions and cylinder length up to $L=22$. Sample calculations
are given in Fig.~(\ref{bubble}). The top panel (a) shows $g({\bf r})$ for the pure $V_1$ model where
we know that the physics of the Laughlin state is the correct one. The probe electron has been set
at the origin of coordinates ($g({\bf r})=G({\bf r},{\bf 0})$) and there is a visible strong correlation hole around it. Beyond a ring of overdensity there is only a featureless fluid. Since we use a truncation of the number of
orbitals there is a range of $x$ coordinates beyond which the density goes to zero~: these are 
the dark boundaries in all panels. If we now increase $V_3$ we observe the appearance of density modulations that are like one-dimensional stripes in panel (b) and (c). Finally when tuning to
the pure $V_3$ model in panel (d) we observe a two-dimensional pattern of density modulation.
If we count the number of peaks we find that each overdensity contains exactly two electrons.
This is what we expect from a bubble phase as predicted for filling factors around $\nu=1/3$
by Hartree-Fock calculations~\cite{Koulakov96,Fogler96,Fogler97}. It has long been observed that weakening the hard-core component
$V_1$ destroys incompressibility in the case of the Coulomb potential. Our results on a truncated model suggests that the topological order of fractional quantum Hall liquids is replaced by the simpler correlations of Hartee-Fock theory such as the bubble phase by weakening $V_1$. We note that extensive ED~\cite{hardcore-us}
in various geometries are in agreement with the compressibility of the pure $V_3$ model.

\section{Studying Coulomb interacting electrons at small filling factor}
\label{CoulombRes}
We now discuss fractional quantum Hall states with the Coulomb interaction.
First of all we check the ground state energy obtained by cylinder DMRG.
The energy per particle of the $\nu=1/3$ liquid in presence of Coulomb interaction (and neutralizing background) is known to be $e_0 \simeq -0.4101$ in the thermodynamic limit. The value $e_0=-0.41016(2)$ has been obtained by DMRG calculations in the spherical geometry with  up to 24
electrons~\cite{zhao_fractional_2011},
and iDMRG calculations in the infinite cylinder geometry \cite{zaletel_infinite_2015} with perimeter $L\leq 20$ found $e_0=-0.410164(4)$.
Here, as a check of our DMRG calculations, we estimate this energy using finite cylinders, both with the Yukawa and chord regularizations. Our results are consistent with the value $e_0 \simeq -0.4101$ previously reported.

\subsection{Energy with Coulomb-chord}

\begin{figure}
\begin{center}
\includegraphics[width=\columnwidth]{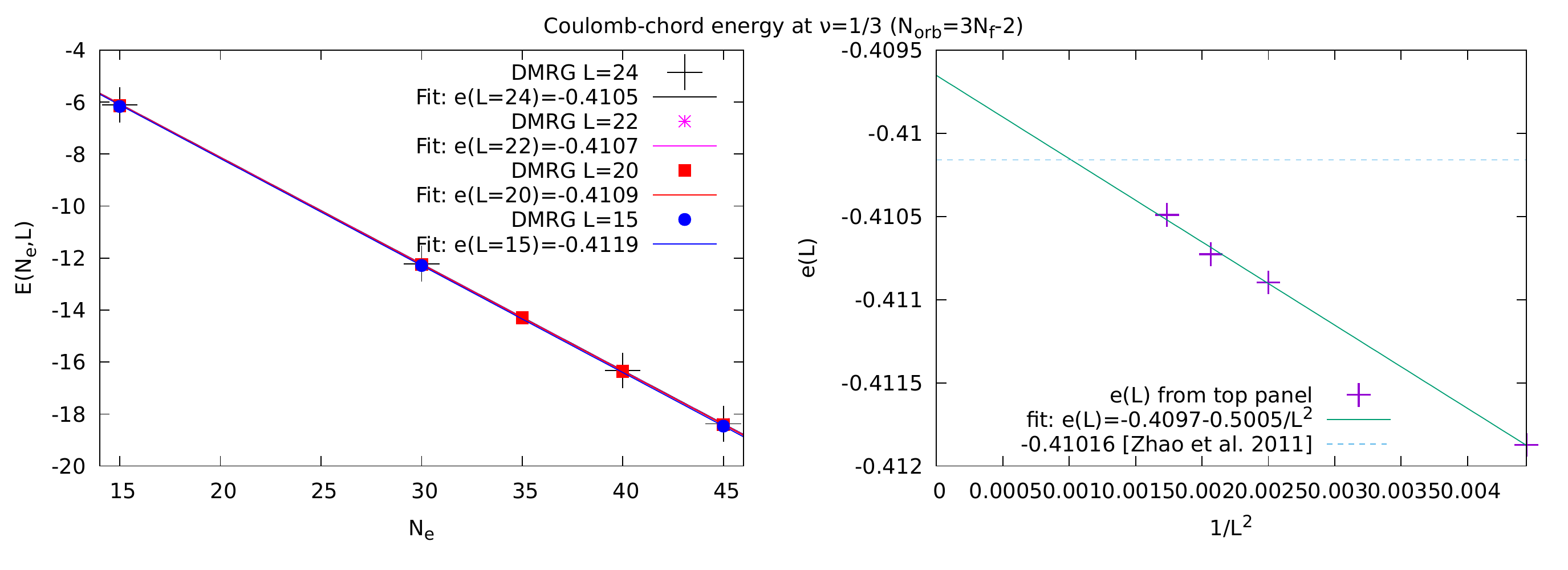}
\caption{Energy at $\nu=1/3$ for the Coulomb-chord interaction potential. Left panel: energy as a
function of $N_e$ for a few values of the circumference $L$ from 15 to 24. In all calculations the
number of orbitals
is $N_{\rm orb}=N_\phi+1=3N_e-2$. These energies
are well fitted by : $E(N_e,L)\simeq e(L) N_e + b(L)$, where $b$ can be interpreted as a
contribution coming from the edge regions of the cylinder.
Right panel: Extrapolation of $e(L)$ (obtained in the left panel) to $L=\infty$. All the calculations have been performed with a maximum bond dimension of the order of 8000.
} 
\label{fig:e_chord}
\end{center}
\end{figure}

As shown in the upper panel of Fig.~(\ref{fig:e_chord}), the energy $E(N_e,L)$ is approximately
linear in the number of particles (and thus also linear in the length $L_x$ of the cylinder), which
allows to extract an energy per particle $e(L)$.  These energies $e(L)$ can then be extrapolated to
the limit of large circumference $L$, as illustrated in the bottom panel of 
Fig.~(\ref{fig:e_chord}).
The result of the above extrapolation to $L=\infty$ is $e_0\simeq -0.4097$, which agrees with 
previous results~\cite{zhao_fractional_2011,zaletel_infinite_2015} up to a relative error of
$10^{-3}$.
The reason why finite-circumference corrections to the energy should scale as $\mathcal{O}(L^{-2})$
is simply an effect of the curvature of the cylinder.
If we set the chord distance to be $d=(L/\pi)\sin\left(\pi r /L\right)$ we have
$d/r \simeq 1 + \frac{1}{6}\left(\pi r /L\right)^2+\mathcal{O}\left((r/L)^4\right)$.
So, at distances $r\ll L$ the Coulomb-chord potential differs from $1/r$ by some a 
$\mathcal{O}\left((r/L)^2\right)$ correction.
In the calculation of the energy, the relevant distances are at most of the order of the
correlation length $\xi$ (as measured by the pair correlation function), and
the leading finite-$L$ correction to $e(L)$ are thus expected to be 
$\mathcal{O}\left((\xi/L)^2\right)$.

\subsection{Energy with Coulomb-Yukawa}

\begin{figure}
\begin{center}
\includegraphics[width=\linewidth]{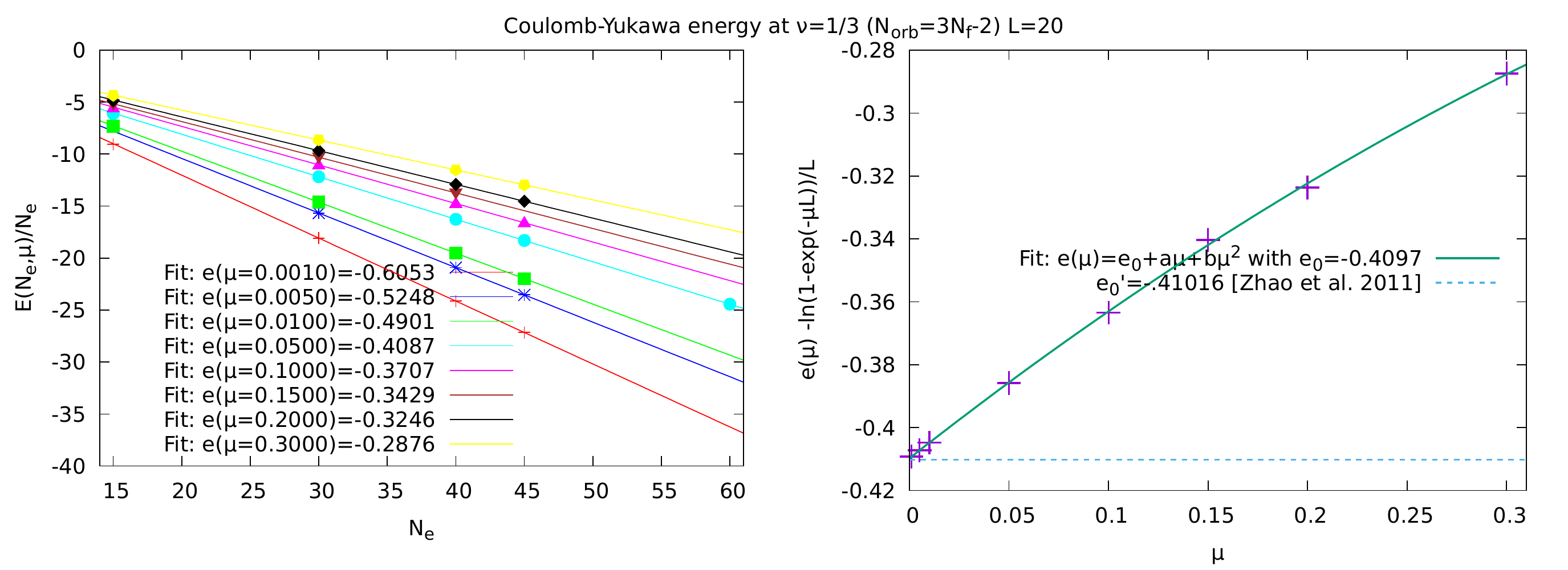}
\caption{Energy at $\nu=1/3$ for the Coulomb-Yukawa interaction potential. Left panel: energy as a
function of $N_e$ for a few values of $\mu$ and fixed $L=20$.
}
\label{fig:e_yukawa}
\end{center}
\end{figure}

The energy data at $\nu=1/3$ with the Coulomb-Yukawa interaction ($\exp(-\mu r)/r$) are plotted in Fig.~\ref{fig:e_yukawa}
as a function of $N_e$ for a few values of $\mu$ between $10^{-3}$ and $0.3$. Here the
circumference is fixed to $L=20$.
The upper panel shows that energy $E(N_e,\mu)$ is approximately linear in the number of particles (and thus in $L_x$ too), which allows to extract an energy per particle $e(\mu)$.
So far the energy takes into account the electron-electron interaction, electron-background potential, and background-background energy. What is however missing is the energy due to the Coulomb interaction between each electron and its own periodic images (see Eq.~\ref{eq:coulomb_images}). This contributes to the energy by a constant $e_P$ per particle
which diverges logarithmically when $\mu\to0$~:
\begin{equation}
e_P= \sum_{n=1}^\infty \frac{e^{-\mu n L}}{n L} = \frac{1}{L} \ln \left(1-e^{-\mu L}\right).
\end{equation}
Adding $e_P$ to the extrapolated DMRG energies gives the data plotted in the bottom panel of Fig.~\ref{fig:e_yukawa}. The corrected energy is extrapolated to $e_0\simeq-0.4097$ in the limit $\mu\to0$. This is again in agreement with the previous estimates.

\subsection{Orbital occupancies}
\begin{figure}
\begin{center}
\includegraphics[width=0.8\columnwidth]{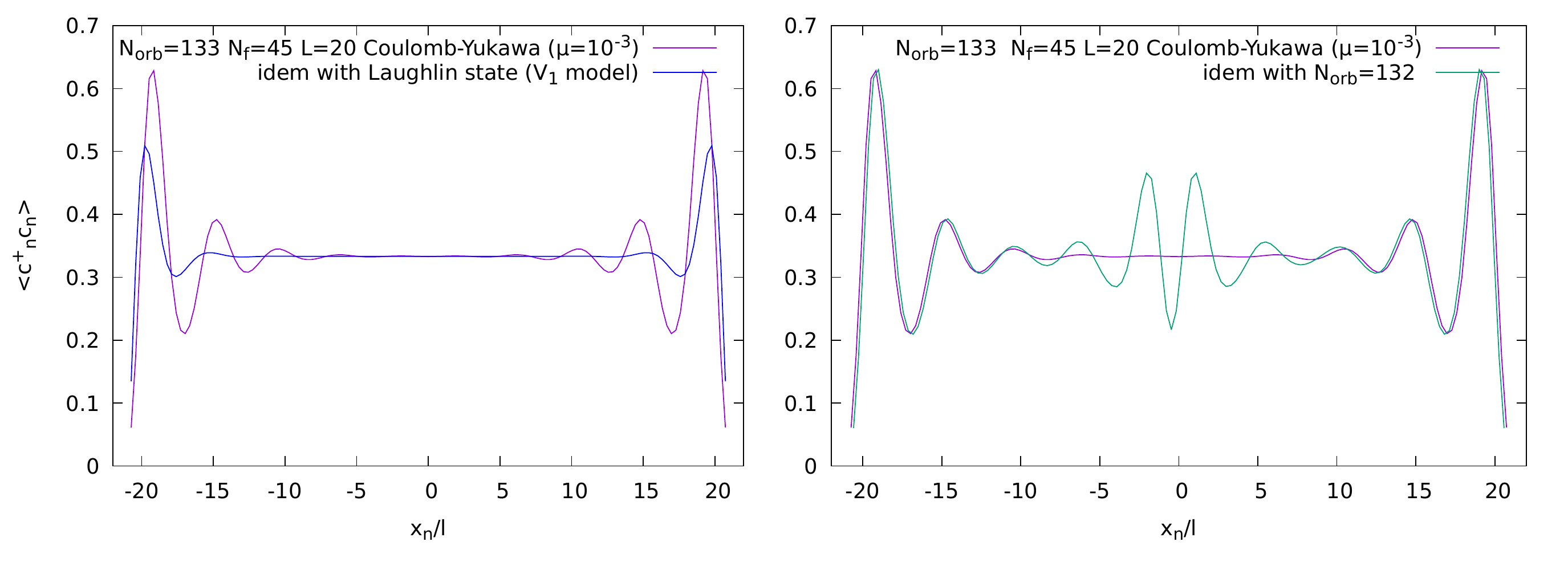}
\caption{Left: Laughlin state occupation numbers in blue compared to Coulomb ground state
from Yukawa potential with $\mu=10^{-3}$. Right: creation of one quasielectron by removal of one
orbital state.
}
\label{occnumbers}
\end{center}
\end{figure}

With faith in the convergence of DMRG we now investigate the physical properties of the incompressible
states. In Fig.~(\ref{occnumbers}) we display the occupation numbers of the Laughlin state obtained by finding
the ground state of the $V_1$ model and the ground state of the Coulomb-Yukawa model
for the same number of electrons/orbitals. We see that there are strong boundary effects near the 
physical end of the cylinder but they are quickly damped when one enters the bulk of the system.
There is a wide region with uniform occupation that should capture the bulk physics at this filling factor. When we increase the number particles we observe that the regions of oscillating behavior
are more and more far apart, increasing the range of the region whose behavior is that of the bulk.
In the right panel of Fig.~(\ref{occnumbers}) we have removed one orbital from the cylinder, an operation that creates
a quasielectron on top of the $\nu=1/3$ liquid. Indeed we observe in the center of the cylinder
the appearance of a crater-like density modulation which is consistent with what we know about such
quasiparticles. As a reference we have also plotted the occupation for the fiducial Laughlin state.
The quasielectron is indeed localized in real space with an extent of the order of a few
magnetic lengths.

\subsection{Wigner crystal}
\begin{figure}
\begin{center}
\includegraphics[width=\linewidth]{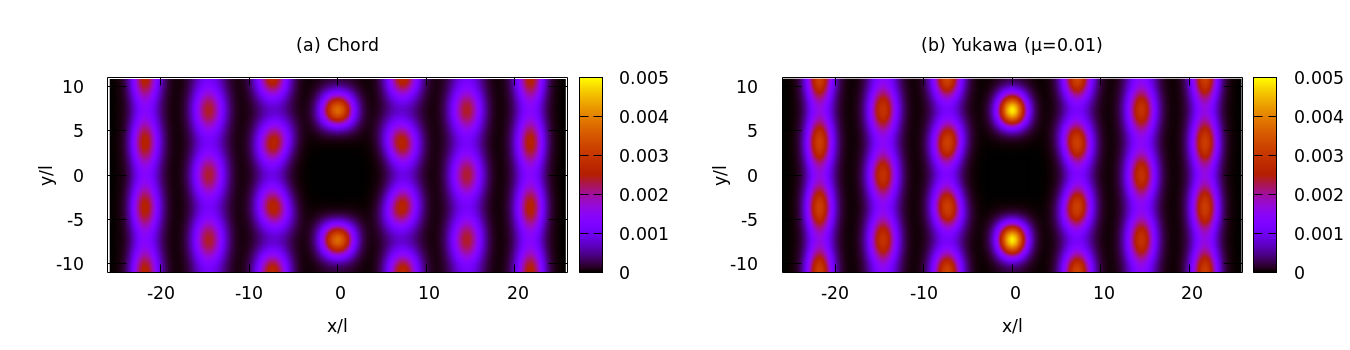}
\caption{Pair correlation at $\nu=1/9$ for $N_e=21$ electrons and a cylinder with $L=22$.
left panel is obtained with the chord distance regularization while the right panel
is from Yukawa potential with $\mu=0.01$. The cylinder extent has been fine-tuned to maximize
the appearance of the Wigner crystal state.
In the chord and Yukawa cases the largest truncation weights are  $\sim 3.10^{-9}$ and $\sim 10^{-9}$, for largest bond dimensions equal to $\sim 6900$ and $\sim 7600$ respectively.
}
\label{WXtal}
\end{center}
\end{figure}

We now turn to the fate of the electron system at very small filling
 factors~\cite{zuo_how_2020}.
It is known that the FQHE liquids compete with the electron solid called the Wigner crystal
and at small filling factors the Wigner crystal is expected to become the ground state of the 
system.
The pair correlation function is well suited to reveal the crystal state since it displays directly 
the spatial density modulation as we have seen in the example of the hard-core model
in section (\ref{hard}). We have performed calculations for filling factor $\nu=1/9$ using a number 
of particles of up to $N_e=21$. To accommodate a triangular crystal in a finite cylinder it is more favorable to
have a number of particles which is a multiple of 3. With the truncation in space along the $x$ 
direction and the finite extent in the $y$ direction
on should adjust $L$ so that the aspect ratio of the cylinder does not frustrate the expected triangular-lattice pattern,
and so that a clean crystal structure can be observed. Such a case is displayed for $N_e=21$ electrons
in Fig.~(\ref{WXtal}) where we show that use of chord or Yukawa interactions
leads to the same kind of crystal state.

\section{Conclusions}
\label{conclude}

Numerical techniques have been an invaluable tool in the understanding of the fractional quantum
Hall effect. Even when severely size-limited they give an unbiased information about the ground 
state correlations and the nature of elementary excitations. The size limitation becomes
problematic in several areas where rapid experimental progress challenge theoretical
understanding. Notably this includes quantum Hall fractions with complicated commensurabilities
like $4/11$ or $6/13$. This also severely hampers the understanding of multicomponent systems
like graphene that involve a fourfold degeneracy of Landau levels due to spin and valley degrees 
of freedom. Multilayer systems made by stacking monolayer materials or fabricating
on-purpose special devices are also a frontier where advances are needed. By looking
at the Landau level problem from a one-dimensional point of view  it is natural to use the DMRG
algorithm with the obvious obstacle that interactions remain long-range in the physical regime.
Indeed in the Hamiltonian Eq.(\ref{pseudoH}) the decay factor governed by 
$\lambda=\exp(-2\pi^2/L^2)$ is less and less effective as a cut-off on the range of interactions 
as we go to larger cylinders $L\rightarrow\infty$. When the range of interactions is large the 
so-called MPO representation of the Hamilotnian grows in complexity and the MPS bond dimensions
also grows. So there is a trade-off also in DMRG involving all these parameters. 
We have presented a set of problems that have been studied following this path.
We have seen
that the present-day technology is advanced enough so that previously intractable problems
in the FQHE realm can now be studied. For example there is no obvious obstacle to include the
multicomponent nature of electrons in the near future.

\ack We thank E. Burovsky  and CSP organizers for invitation at CSP2020. 
One of us (ThJ) thanks Song-Yang Pu and Zheng-Wei Zuo for useful discussions.
We also 
acknowledge CEA-DRF for computer time allocation on the supercomputer
COBALT at CCRT.

\section*{References}


\begin{thebibliography}{10}
\expandafter\ifx\csname url\endcsname\relax
  \def\url#1{{\tt #1}}\fi
\expandafter\ifx\csname urlprefix\endcsname\relax\def\urlprefix{URL }\fi
\providecommand{\eprint}[2][]{\url{#2}}

\bibitem{white_density_1992}
White S~R 1992 {\em Phys. Rev. Lett.\/} {\bf 69} 2863--2866
  \urlprefix\url{http://link.aps.org/doi/10.1103/PhysRevLett.69.2863}

\bibitem{SchollwockAnnPhys11}
Schollw{\"o}ck U 2011 {\em Annals of Physics\/} {\bf 326} 96 -- 192 ISSN
  0003-4916
  \urlprefix\url{http://www.sciencedirect.com/science/article/pii/S0003491610001752}

\bibitem{itensor3}
Fishman M, White S~R and Stoudenmire E~M 2020 {\em arXiv:2007.14822 [cond-mat,
  physics:physics]\/} ArXiv: 2007.14822
  \urlprefix\url{http://arxiv.org/abs/2007.14822}

\bibitem{shibata_ground-state_2001}
Shibata N and Yoshioka D 2001 {\em Phys. Rev. Lett.\/} {\bf 86} 5755--5758
  \urlprefix\url{https://link.aps.org/doi/10.1103/PhysRevLett.86.5755}

\bibitem{yoshioka_dmrg_2002}
Yoshioka D and Shibata N 2002 {\em Physica E: Low-dimensional Systems and
  Nanostructures\/} {\bf 12} 43--45 ISSN 1386-9477
  \urlprefix\url{http://www.sciencedirect.com/science/article/pii/S138694770100306X}

\bibitem{shibata_ground_2003}
Shibata N and Yoshioka D 2003 {\em J. Phys. Soc. Jpn.\/} {\bf 72} 664--672 ISSN
  0031-9015 \urlprefix\url{https://journals.jps.jp/doi/10.1143/JPSJ.72.664}

\bibitem{feiguin_density_2008}
Feiguin A~E, Rezayi E, Nayak C and Das~Sarma S 2008 {\em Phys. Rev. Lett.\/}
  {\bf 100} 166803
  \urlprefix\url{https://link.aps.org/doi/10.1103/PhysRevLett.100.166803}

\bibitem{kovrizhin_density_2010}
Kovrizhin D~L 2010 {\em Phys. Rev. B\/} {\bf 81} 125130
  \urlprefix\url{https://link.aps.org/doi/10.1103/PhysRevB.81.125130}

\bibitem{zhao_fractional_2011}
Zhao J, Sheng D~N and Haldane F~D~M 2011 {\em Phys. Rev. B\/} {\bf 83} 195135
  \urlprefix\url{https://link.aps.org/doi/10.1103/PhysRevB.83.195135}

\bibitem{zaletel_infinite_2015}
Zaletel M~P, Mong R~S~K, Pollmann F and Rezayi E~H 2015 {\em Phys. Rev. B\/}
  {\bf 91} 045115
  \urlprefix\url{https://link.aps.org/doi/10.1103/PhysRevB.91.045115}

\bibitem{zhu_fractional_2015}
Zhu W, Gong S, Haldane F and Sheng D 2015 {\em Phys. Rev. Lett.\/} {\bf 115}
  126805
  \urlprefix\url{https://link.aps.org/doi/10.1103/PhysRevLett.115.126805}

\bibitem{zhu_topological_2015}
Zhu W, Gong S~S, Haldane F~D~M and Sheng D~N 2015 {\em Phys. Rev. B\/} {\bf 92}
  165106 \urlprefix\url{https://link.aps.org/doi/10.1103/PhysRevB.92.165106}

\bibitem{johri_probing_2016}
Johri S, Papic Z, Schmitteckert P, Bhatt R~N and Haldane F~D~M 2016 {\em New J.
  Phys.\/} {\bf 18} 025011 ISSN 1367-2630
  \urlprefix\url{https://doi.org/10.1088%2F1367-2630%2F18%2F2%2F025011}

\bibitem{zhu_fractional_2016}
Zhu W, Liu Z, Haldane F~D~M and Sheng D~N 2016 {\em Phys. Rev. B\/} {\bf 94}
  245147 \urlprefix\url{https://link.aps.org/doi/10.1103/PhysRevB.94.245147}

\bibitem{Soule1}
Soulé P and Jolicoeur T 2012 {\em Phys. Rev. B\/} {\bf 85} 155116 publisher:
  American Physical Society
  \urlprefix\url{https://link.aps.org/doi/10.1103/PhysRevB.85.155116}

\bibitem{Soule2}
Soulé P and Jolicoeur T 2012 {\em Phys. Rev. B\/} {\bf 86} 115214 publisher:
  American Physical Society
  \urlprefix\url{https://link.aps.org/doi/10.1103/PhysRevB.86.115214}

\bibitem{Soule3}
Soulé P, Jolicoeur T and Lecheminant P 2013 {\em Phys. Rev. B\/} {\bf 88}
  235107 publisher: American Physical Society
  \urlprefix\url{https://link.aps.org/doi/10.1103/PhysRevB.88.235107}

\bibitem{Wojs2004}
Wójs A, Yi K~S and Quinn J~J 2004 {\em Phys. Rev. B\/} {\bf 69} 205322
  publisher: American Physical Society
  \urlprefix\url{https://link.aps.org/doi/10.1103/PhysRevB.69.205322}

\bibitem{Koulakov96}
Koulakov A~A, Fogler M~M and Shklovskii B~I 1996 {\em Phys. Rev. Lett.\/} {\bf
  76} 499--502 publisher: American Physical Society
  \urlprefix\url{https://link.aps.org/doi/10.1103/PhysRevLett.76.499}

\bibitem{Fogler96}
Fogler M~M, Koulakov A~A and Shklovskii B~I 1996 {\em Phys. Rev. B\/} {\bf 54}
  1853--1871 publisher: American Physical Society
  \urlprefix\url{https://link.aps.org/doi/10.1103/PhysRevB.54.1853}

\bibitem{Fogler97}
Fogler M~M and Koulakov A~A 1997 {\em Phys. Rev. B\/} {\bf 55} 9326--9329
  publisher: American Physical Society
  \urlprefix\url{https://link.aps.org/doi/10.1103/PhysRevB.55.9326}

\bibitem{hardcore-us}
Misguich G, Jolicoeur Th and Mizusaki T 2020 {\em arXiv:1703.07095 [cond-mat]\/}
  ArXiv: 1703.07095 \urlprefix\url{http://arxiv.org/abs/1703.07095}

\bibitem{zuo_how_2020}
Zuo Z~W, Balram A~C, Pu S, Zhao J, Jolicoeur Th, Wójs A and Jain J~K 2020 {\em
  Phys. Rev. B\/} {\bf 102} 075307 publisher: American Physical Society
  \urlprefix\url{https://link.aps.org/doi/10.1103/PhysRevB.102.075307}

\end{thebibliography}

\providecommand{\newblock}{}


\end{document}